\documentstyle[aps,twocolumn,epsf,rotate,prl]{revtex}
\newcommand{\be}{\begin{equation}}
\newcommand{\ee}{\end{equation}}
\newcommand{\bdm}{\begin{displaymath}}
\newcommand{\edm}{\end{displaymath}}
\newcommand{\ba}{\begin{eqnarray}}
\newcommand{\ea}{\end{eqnarray}}

\begin{document}

\title{
 Magnetic Field Effect on the Supercurrent of an SNS junction 
} 
\author{S.-K. Yip}
\address{
Physics Division, National Center for Theoretical Sciences ,
P. O. Box 2-131,  Hsinchu, Taiwan 300,
R. O. C.}
\date{\today}

\maketitle

\begin{abstract}
{\small

In this paper we study the effect of a Zeeman field
on the supercurrent of a mesoscopic SNS junction.
It is shown that the supercurrent suppression is
due to a redistribution of current-carrying
states in energy space.  A dramatic consequence is
that (part of the) the suppressed supercurrent can be recovered
with
a suitable non-equilibrium distribution of
quasiparticles.

 }
\end{abstract}
\date{\today}
\vspace*{0.2 cm}

Recently, there has been a revival of interest in the proximity
effect \cite{deGennes} which occurs when normal metals (N) are
placed in contact with superconductors (S).
This is due to the availability of submicron fabrication technology 
 and low temperatures.  One of
the main lessons learnt in the study of these hybrid structures
is that, at sufficiently low temperatures, due to
the suppression of inelastic scattering etc.,
it is essential to properly understand the contribution
from individual energies. 
\cite{Beenakker97,Lambert97,Courtois99,SM99}.
 This applies to, e.g.,
the conductance between N and S in contact
as well as supercurrent in an SNS junction
\cite{Morpurgo98,Baselmans99,Yip98,Wilhelm98}

For the latter systems, it is in particular useful
to understand the supercurrent being carried at
each individual energy, i.e., the
current-carrying density of states $N_J(\epsilon)$.
This quantity is the ordinary density of states
weighted by the current that the states carry.
The supercurrent is the integral over 
the energy $\epsilon$ of the product
of $N_J(\epsilon)$ and the occupation number $n(\epsilon)$.

Here we consider again an SNS junction but with
an applied magnetic field.   It is well-known that
magnetic field in general suppresses the supercurrent.
This can arise from two completely different
mechanisms \cite{Tinkham}.  First, it can be due to
the coupling of the magnetic field to superconductivity
via the vector potential.  I shall not
discuss this effect here.  This effect can be suppressed
in geometries where the area perpendicular to
the magnetic field is sufficiently small.  The
second mechanism is due to the Zeeman energy.
Since pairing is singlet in s-wave superconductors, 
a physical picture
 commonly used is that the Zeeman field is
a pair-breaking perturbation and hence 
the Cooper pair amplitude
decays in space faster in the presence of the field.
In a dirty normal metal with diffusion coefficient $D$, this decay length 
$\sim \sqrt{D/h}$ \cite{Tinkham} for an energy splitting $h$.
For an SNS junction, the supercurrent
thus decreases with field due to a reduction of coupling
between the two superconductors.

This, however, is not entirely the full picture.
As will be shown below, the main effect of the Zeeman
splitting is to shift the current-carrying density of
states in energy space. 
This is analogous to the behavior of the 
ordinary density of states under $h$ \cite{Tinkham}.
 The pairing correlation between
the two superconductors remains {\it long-ranged} at
appropriate energies.
The supercurrent decreases ( {\it c.f.}, however, below)
because of this mentioned shift and the associated change
in the occupation of the states (see below for details).

A dramatic consequence of the above is that,
under a suitable non-equilibrium distribution
of quasiparticles, one can {\it recover}
the suppressed supercurrent.  I shall demonstrate
this using an experimental arrangement studied
in ref \cite{Morpurgo98,Baselmans99}.  A suitable
applied voltage can recover (partly) the 
supercurrent suppressed by a magnetic field.

Consider thus a quasi-one dimensional dirty metal wire
(N') of length $L$ 
connecting two superconductors.   We shall always
assume that the junction is in the  dirty limit.
The supercurrent carrying density of states can
be obtained via the angular averaged retarded green's function 
matrix $\hat g$.
For the present purposes  
 it is sufficient to
 consider the retarded component and I shall leave out
 the usual superscript $R$.
 $\hat g$  obeys the normalization condition
$\hat g ^2 = - \pi^2 \hat 1 \ $ and the Usadel equation.
The latter reads, for position $x$ within N' ($ 0 < x < L$) and
  a magnetic field $B$ along the $\hat z$ direction, 

\begin{equation}
[\epsilon \hat \tau_3 + h \hat \sigma_3, \hat g] + 
 {D \over \pi} \partial_x (\hat g  \partial_x \hat g) = 0
\label{usadel}
\end{equation}

\noindent Here
$\epsilon$ is the energy with respect to the Fermi level, 
and $h = \mu_e B$.  In order to avoid confusions
I shall pretend that electrons have a positive
magnetic moment and thus identify the 
directions of the magnetic moment and spin,
with up (down) being the states with lower (higher) Zeeman energy.
$\hat g$ at the boundaries $ x = 0 $ and $L$  are given by its 
corresponding values for the equilibrium superconductors.
I shall assume that the magnetic field is perfectly
screened in S.  In this case  
the boundary conditions at the superconductors are given
by 
\be
\hat g = - \pi {  \epsilon \hat \tau_3 - \hat \Delta 
\over \sqrt{ |\Delta|^2 - \epsilon^2 }}.
\label{bc}
\ee

\noindent with suitable gap matrices $\hat \Delta$
reflecting the phase difference
$\chi$  between the two superconductors.

Eqn (\ref{usadel}) can be simplified by noting that
it is block-diagonal, since the pairing
is singlet.  In the usual $4 \times 4$
notation, the elements associated with
the 1st and 4th rows and columns are decoupled from those
of 2nd and 3rd 
 (as already noted in, {\it e.g.}, ref \cite{Buzdin,Demler97})
Moreover, the matrix equations for these submatrices have
the same structure as that in zero field except
$\epsilon \rightarrow \epsilon \pm h$,
corresponding to magnetic moment parallel and antiparallel
with the external field.  It is then convenient to introduce
separately the current-carrying density of states for
each spin direction:

\be
N^{\sigma}_J (\epsilon) =  < \hat p_x \ N^{\sigma} (\hat p, \epsilon, x) >
\label{defNJ}
\ee

\noindent where $N^{\sigma} (\hat p, \epsilon, x)$ is the 
density of states for spin $\sigma$ ( $ = \ \uparrow $ or $\downarrow$),
$\hat p$ the direction of momentum.
Here I have chosen to label the states with the spin
direction of the particles.  Note that, {\it e.g.},
the up-spin particles are associated with the down-spin holes
({\it c.f.} above).
For a given spin $\sigma$, $N^{\sigma}_J (\epsilon)$ is
related to the appropriate sub-matrix of the green's function
$\hat g$ via formulas  analogous to
those in zero field  (see, e.g. \cite{Yip98})

The total (number) current density at $T= 0$ is given by
the integration of $N_J^{\uparrow \downarrow}$
 over the occupied (negative) energy
states:

\be
J_s =  v_f \int_{-\infty}^0 {d \epsilon} \ 
[ N^{\uparrow} _J ( \epsilon) +  N^{\downarrow} _J ( \epsilon) ]
\label{Js0}
\ee

We shall also introduce
$N^{av}_J (\epsilon) = {1 \over 2} 
[  N^{\uparrow} _J ( \epsilon) +  N^{\downarrow} _J ( \epsilon) ] $
as the spin averaged current-carrying density of states.
$J_s$ is related to $N^{av}_J(\epsilon)$ by
\be
J_s =  2 v_f \int_{-\infty}^0 {d \epsilon} \ 
[ N^{av} _J ( \epsilon)  ], 
\label{Jsav}
\ee

\noindent exactly
the same formula as in zero field. Here
$v_f$ is the Fermi velocity.

\begin{figure}
\centerline{
        \epsfysize=0.4\textwidth \rotate[r]{
        \epsfbox{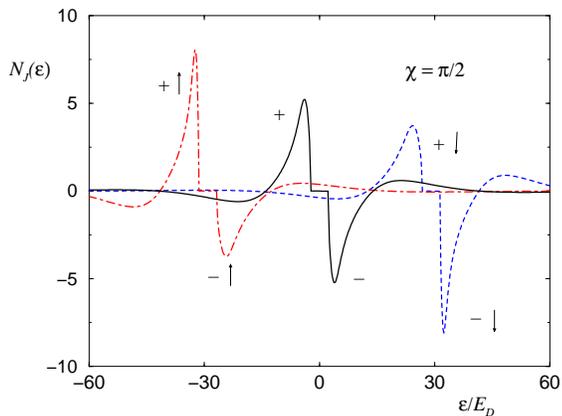} }
}
\vskip 0.5 cm
\begin{minipage}{0.45\textwidth}
\caption[]{
$N^{\uparrow \downarrow}_J$ 
(dot-dashed and dashed respectively and in units of $N_f l/6 L$
\cite{unit})
 for a junction with $\Delta = 100 E_D$ at $h = 30 E_D$.
 Also shown is $N_J (\epsilon) $ in zero field (full line).
 }
\label{fig:njh30}
\end{minipage}

\end{figure}

We shall confine ourselves to the case of long
junctions ( $ E_D << | \Delta | $ ),
and for definiteness choose $|\Delta| = 100 E_D$.  Here
$E_D \equiv D/L^2$ is the Thouless energy
associated with N'.  The behavior of
the current-carrying density of states $N_J(\epsilon)$
in zero field
has already been studied in detail in \cite{Yip98,Wilhelm98}.
I shall only mention some of the more relevant features below.
$N_J(\epsilon)$ vanishes for all energies at $\chi = 0$.  
A typical case for other phase differences $ 0 < \chi < \pi$
(we shall always restrict ourselves to this range,
the other cases can be obtained by symmetries)
 is as shown by
full line in Fig \ref{fig:njh30}.  
$N_J$ is odd in the energy variable $\epsilon$. Its major feature
consists of a
positive peak (labelled by $+$ in Fig \ref{fig:njh30})
at energies of several times 
the Thouless energy $E_D$ below the fermi energy,
 and a corresponding negative
peak above (labelled by $-$ in Fig. \ref{fig:njh30}).
  Also seen are the small undulations
as a function of energy for larger energies.  This oscillatory
behavior is a result
of the difference in wave-vectors for the participating
particles and holes undergoing Andreev reflection at given $\epsilon$.
For one dimension and in the clean case
the pairing amplitude $f$ 
(the off-diagonal elements of $\hat g$ in particle-hole space)
oscillates as $e^{\pm 2 i ( \epsilon / v_f ) x}$.
In the present dirty three dimensional case the same
physics results in {\it e.g.},
$f \sim exp \pm [( 1 - i ) \sqrt{ \epsilon / D } x ] $
for the linearized Usadel equation ({\it i.e.}
the limit of small pairing amplitudes) 
At a fixed position, the pairing amplitude and hence the
coupling between the two superconductors oscillates
as a function of energy. 
$N_J(\epsilon)$ also vanishes for $|\epsilon|$ below a few
times $E_D$, where the ordinary density of states 
also vanishes.  Both the magnitude of this `minigap' and
the position of peaks mentioned above decrease with increasing
phase difference, vanishing as $\chi \to \pi$
(where $N_J$ itself also vanishes for all $\epsilon$'s).

The behavior of $N^{\sigma}_J(\epsilon)$ under a finite field is
also shown in Fig \ref{fig:njh30}, where I have
chosen an intermediate $h$ ( $ E_D << h << |\Delta|$) 
for clarity. As mentioned it is convenient to
discuss the current-carrying density of states
separately for each spin direction
under the presence of $h$.
For magnetic moment along the applied field, 
the current carrying density of states is roughly ({\it c.f.} below)
that of zero field except shifted in energy by $-h$.
{\it i.e.} $N_J^{\uparrow} (\epsilon) 
\approx N_J^{h = 0}(\epsilon + h)$.
Correspondingly $N^{\downarrow}_J$ for magnetic moment pointing in
the opposite direction is shifted up in energy.
At fields $h$ not too small compared with $\Delta$,
there is a correction to this picture because,
if we assume perfect screening of the magnetic field
inside the superconductor as we are doing,
the replacement 
 $\epsilon \rightarrow \epsilon \pm h$
 in the Usadel equation does not apply
  for the boundary condition (\ref{bc})
at $ x = 0$ and $L$.  This correction is negligible if
$ h << |\Delta|$ and increases with increasing $h$.
 
 In this picture, the reason that the supercurrent
 at finite $h$ is suppressed (in general) from that
 of zero field is {\it not} pair-breaking, at least
 for $h << | \Delta|$.  Rather,
 it is because at finite field $\sim E_D$, some of the states
 that have positive contributions to $J_s$ were orginally
 occupied at $h = 0$ but are now empty ($ + \downarrow$).,
 whereas some which were orginally empty are now
 occupied ( $- \uparrow$) and contribute a negative current.

\begin{figure}
\centerline{
        \epsfysize=0.4\textwidth \rotate[r]{
        \epsfbox{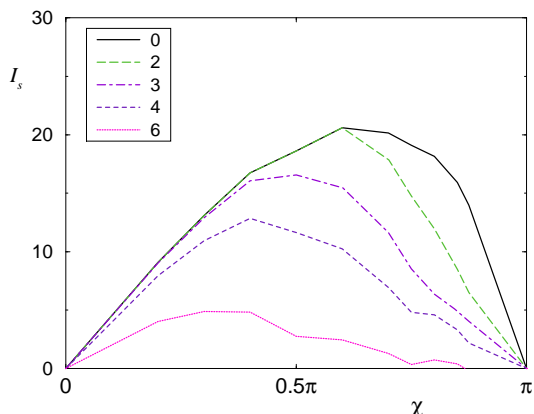} }
}
\vskip 0.5 cm
\begin{minipage}{0.45\textwidth}
\caption{
Current-phase relationships for magnetic
fields $h$ displayed in the legend.  $h$ is
in units of $E_D$,
$I_s$
  in units of $ E_D/ 2 R_N$. 
Here  $R_N$ is the normal state resistance
 of N' ($ {1 \over R_N} = 2 N_f D S L^{-1} $
 with $S$ the cross-section area of N').}
\label{fig:ichi}
\end{minipage}

\end{figure}

 By the above reasoning,
 the presence of the magnetic field has a non-trivial
 effect on the current-phase relationship.  
 An example for a small $h << |\Delta|$ is
 shown in Fig \ref{fig:ichi}.  In zero field
 $I(\chi)$ is roughly like a sine function except
 for a small tilt towards $\chi \approx \pi$. 
 When $h$ increases from zero, one sees that $I_s$ 
 first starts to decrease for $\chi$ near $\pi$
 while $I_s$ at 
 smaller $\chi$ is unaffected.   Only at larger
 $h$ would $I_s$ begin to be suppressed there.
 This can be readily understood by considering
 the behavior of $N_J$ under $h$ discussed before. 
 Recall that at zero field $N_J$ has a 
minigap of order $E_D$ but $\chi$ dependent, being smallest
when $\chi$ is near $\pi$. 
 Thus when the field $h$
is increased from $0$, $I_s$ at larger 
$\chi$ would be suppressed first 
since at these $\chi$'s,  a smaller $h$ is needed
to shift the antipeak $- \uparrow$ (the peak $ + \downarrow$) 
to below (above) the fermi level.

At higher $h$ and near $\chi \approx \pi$ the current
also oscillates with $\chi$. (See, in particular
$h = 6$ in Fig. \ref{fig:ichi}, where $I_s$ becomes negative
for $\chi$ slightly less than $\pi$, vanishing
again at $\chi = \pi$).  These features
are due to the undulatory structure of $N_J^{\sigma}$
as a function of $\epsilon$ (see Fig \ref{fig:njh30}).
  At these higher fields
the weaker bumps and troughs of $N^{\sigma}_J$
(not labelled) cross the fermi level successively.
They do so at  fields which are $\chi$ dependent.
Their amplitudes also depends on $\chi$.

\begin{figure}
\centerline{
        \epsfysize=0.4\textwidth \rotate[r]{
        \epsfbox{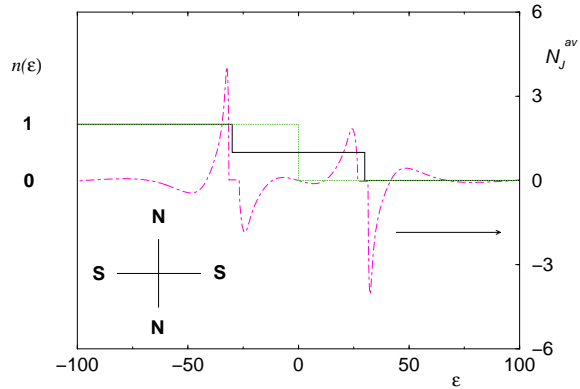} }
}
\vskip 0.5 cm
\begin{minipage}{0.45\textwidth}
\caption{ Explanation of the effect of a voltage
on the supercurrent for the device shown in the inset.
The occupation numbers $n(\epsilon)$ (left scale) are shown
for $V= 0$ (dotted) and $V = 30$ (full line).
Also shown is the spin-averaged current-carrying density
of states $N^{av}_J$ (right scale).
$\epsilon$ and $e V$ are in units of $E_D$.
}
\label{fig:ex}
\end{minipage}

\end{figure}

\begin{figure}
\centerline{
        \epsfysize=0.4\textwidth \rotate[r]{
        \epsfbox{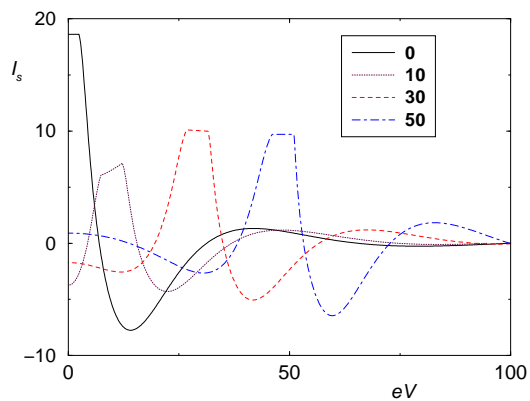} }
}
\vskip 0.5 cm
\begin{minipage}{0.45\textwidth}
\caption{Supercurrent as a function of voltage $V$
between the two superconductors S
of the device depicted in the inset of Fig \ref{fig:ex}.
The values of $h$, in units of $E_D$, are
shown in the legend. $\chi = \pi/2$.}
\label{fig:jhv}
\end{minipage}

\end{figure}

Since the major effect of the magnetic field is 
not a suppression of $N^{\sigma}_J$
 (except for $h { > \atop \sim} |\Delta|$) but
a  redistribution in energy space, the supercurrent
suppression by the magnetic field can, to a certain
extent, be {\it recovered} by a suitable distribution
of quasiparticles.  Here we consider the 
case of a `controllable Josephson Junction',
studied in \cite{Morpurgo98,Baselmans99} experimentally
and \cite{Yip98,Wilhelm98} theoretically.
The device configuration is shown schematically
in the inset of Fig  \ref{fig:ex}.
The two superconductors S, in general with a phase
difference $\chi$, are at zero voltage.
Equal but opposite voltages $V$ are applied on the
normal (N) reservoirs. 
The N and S reservoirs are connected by quasi-one dimensional
normal wires as shown.  We are interested in
the effect of the voltage $V$ on the supercurrent $I_s$ between
the S reservoirs.
 At $V= 0$, the distribution
function  $n(\epsilon)$ is of the usual equilibrium form
and is given by $1$ for $\epsilon < 0$ and $0$ otherwise
($T=0$), as shown by dotted lines in Fig \ref{fig:ex}.
Under a finite $V$, $n(\epsilon)$
for $ | \epsilon | <  eV$  becomes $ 1/2$
\cite{half} (full line in Fig \ref{fig:ex}),
 {\it i.e.} its effect is to
transfer half of the quasiparticles for $ - eV < \epsilon < 0$
to the region $ 0 < \epsilon < e V$.
In zero field such a non-equilibrium distribution in
general leads to a decrease of the supercurrent
between the S reservoirs (see \cite{Morpurgo98,Baselmans99,Yip98,Wilhelm98}),
since usually this corresponds to decreasing
(increasing)  the occupation of states which
contribute a positive (negative) current 
[ $+$ ($-$)  in Fig \ref{fig:njh30}]
(full line in Fig \ref{fig:jhv}, \cite{cross}; see ref \cite{Yip98,Wilhelm98}
for the oscillations at higher $V$'s)

The effect of $V$ on $I_s$ for finite $h$ is
also shown in Fig \ref{fig:jhv}. At the fields
chosen ($>> E_D$) $I_s$ at zero voltages have essentially
decreased to zero (see also below).  As claimed,
for a given $h$, a suitable choice of $V$ may
`enhance' the supercurrent. 
For  the present case where $ E_D << h << |\Delta|$,
this enhancement is particularly spectacular 
for $eV \approx h$.
These features can
be understood by examining 
 the spin-averaged
current-carrying density of states $N^{av}_J (\epsilon)$,
also shown in Fig \ref{fig:ex}.
$N^{av}_J (\epsilon)$ is odd in energy.
The behavior of $N^{av}_J$ follows directly from 
$N^{\uparrow \downarrow}_J$ in Fig \ref{fig:njh30}.
For a given $h \ne 0$, when $V$ is increased from zero,
we have a
transfer of the particles as described before
from  the states near
the antipeak just below the fermi level ( due to $ - \uparrow $)
 to the peak above (arising from $ + \downarrow $).
 The supercurrent thus increases.
 The strongest enhancement of the supercurrent
occurs at $eV \approx h$,
since the region where this transfer occurs just covers the antipeak
below the fermi level.
For $ E_D << h << |\Delta|$, roughly half
of the supercurrent at zero voltage can be
recovered at $eV \approx h$.
This is because at this voltage, 
the contributions from the region 
$ - eV < \epsilon < e V$ cancel among themselves
and we are left with the integral over states
with $ \epsilon < - eV$.  The integral of
 $N^{av}_J$ over the energy in this region is
roughly half that of zero field.
(compare Fig \ref{fig:ex} and Fig \ref{fig:njh30})

The above statements are not quantitatively precise 
due to:
(i) $N^{\uparrow \downarrow}_J$ are not simple
shifts of $N_J$ in energy and (ii) oscillatory 
structures of $N^{\uparrow \downarrow}_J$, both
mentioned before.

For $V=0$, as can be seen in Fig \ref{fig:jhv},
the supercurrent does not simply decay monotonicially
with increasing $h$ but rather shows a damped oscillation.
Such a behavior has already been pointed out before
\cite{Buzdin,Demler97}, and was explained in terms
of the oscillation of the pair-amplitude with $h$ 
at a given distance (the separation between
the two superconductors).  The present work provides
a slightly different but closely connected perspective.
$I_s$ oscillates with $h$ because $N^{\sigma}_J$ does
so as a function of energy.  The field $h$ shifts
$N^{\sigma}_J$ in energy space.  The undulatory behavior
results when regions of alternating signs of 
$N^{\sigma}_J$ shift through the fermi level.

Some of the physics discussed in this paper, such
as the effect of $h$ on the current-phase relationship,
is applicable beyond the dirty limit.  I shall
however defer these to a future study.

There is recently strong interest in the physics
of superconductors in contact with a ferromagnetic
material.    Many papers have simply modelled 
the ferromagnet F with a Stoner field.  
e.g. \cite{Beenakker95} ({\it c.f.}, however \cite{Mazin99})
Within this model the Stoner field is formally equivalent
to the Zeeman field here.  A much discussed topic
is the effect of this field $h$ on the number of
conduction channels \cite{Beenakker95}.
This effect is important only when $h$ is comparable
to the fermi energy of N' and has not been included in
the present calculations.

Part of the paper was written using the facilities
of the Chinese University of Hong Kong.  I thank
Profs. P. M. Hui and H. Q. Lin for their help.


\begin{thebibliography}{9}

\bibitem{deGennes} P. G. deGennes, Superconductivity of
Metals and Alloys (Benjamin, New York) (1964)



\bibitem{Beenakker97} C. W. J. Beenakker, Rev. Mod. Phys. 
   {\bf 69}, 731 (1997)
   
\bibitem{Lambert97} C. J. Lambert and R. Raimondi,
J. Phys. Cond. Matter, {\bf 10}, 5901 (1997)

\bibitem{Courtois99} H. Courtois {\it et al},
 J. Low Temp. Phys., {\bf 116}, 187 (1999)

\bibitem{SM99}  See also review articles in Superlattices and Microstructures,
{\bf 25}  (1999)

\bibitem{Morpurgo98} A. F. Morpurgo, T. M. Klapwijk
and B. J. van Wees,  Appl. Phys. Lett. {\bf 72}, 966 (1998)

\bibitem{Baselmans99} J. J. A. Baselmans, A. F. Morpurgo, 
B. J. van Wees and T. M. Klapwijk, 
Nature, {\bf 397}, 43 (1999)

\bibitem{Yip98} S.-K. Yip, Phys. Rev. B {\bf 58}, 5803 (1998)

\bibitem{Wilhelm98} F. Wilhelm, G. Sch\"on and A. D. Zaikin,
  Phys. Rev. Lett. {\bf 81}, 1682 (1998)
 
\bibitem{Tinkham} M. Tinkham, Intro. to Superconductivity,
(McGraw Hill, 1996) 
  
\bibitem{Buzdin}  A. I. Buzdin {\it et al}, 
  JETP Lett, {\bf 35} 178 (1982); {\bf 53} 321 (1991)
  
\bibitem{Demler97}  E. A. Demler, G. B. Arnold and M. R. Beasley,
  Phys. Rev. {\bf 55}, 15174 (1997)
  
\bibitem{unit} Here $N_f$ is the ordinary density of
states per spin for the normal state, $l$ 
the mean free path.  In this unit, $N_J$ is just $-Q$ of ref
\cite{Yip98} numerically.

\bibitem{half}  This is only approximately true: see
 ref \cite{Yip98}.  The corrections, studied in ref\cite{Yip98},
 depend on
 the details of the geometry of the `cross' and we shall ignore
 them for simplicity.


  
\bibitem{cross}  For simplicity, I shall ignore the
effect of the normal arm on  $N_J^{\sigma}$.  This effect
has been studied in \cite{Yip98} and is mainly an
overall suppression of their magnitudes.



\bibitem{Beenakker95} M. J. M. de Jong and C. W. J. Beenakker,
  Phys. Rev. Lett. {\bf 74}, 1657 (1995)

\bibitem{Mazin99} I. I. Mazin,
  Phys. Rev. Lett. {\bf 83}, 1427 (1999)

\end{thebibliography}
\end{document}